\documentclass[aps,prl,english,twocolumn,superscriptaddress,floatfix,10pt]{revtex4-1}
\usepackage{hackref}
\usepackage{graphicx}
\usepackage{natbib}
\usepackage{mathtools}
\usepackage{amsmath}
\usepackage{amsfonts}
\usepackage{hyperref}
\hypersetup{hypertexnames=false}
\begin{document}

\title{Stability of quantum degenerate Fermi gases of tilted polar molecules}

\author{Vladimir Velji\'{c}}
\affiliation{Scientific Computing Laboratory, Center for the Study of Complex Systems, Institute of Physics Belgrade, University of Belgrade, Pregrevica 118, 11080 Belgrade, Serbia}
\author{Axel Pelster}
\affiliation{Physics Department and Research Center OPTIMAS, Technical University of Kaiserslautern, Erwin-Schr\"odinger Stra\ss e 46, 67663 Kaiserslautern, Germany}
\author{Antun Bala\v{z}}
\email{antun.balaz@ipb.ac.rs}
\affiliation{Scientific Computing Laboratory, Center for the Study of Complex Systems, Institute of Physics Belgrade, University of Belgrade, Pregrevica 118, 11080 Belgrade, Serbia}

\begin{abstract}
A recent experimental realization of quantum degenerate gas of $^{40}$K$^{87}$Rb molecules opens up prospects of exploring strong dipolar Fermi gases and many-body phenomena arising in that regime. Here we derive a mean-field variational approach based on the Wigner function for the description of ground-state properties of such systems. We show that the stability of dipolar fermions in a general harmonic trap is universal as it only depends on the trap aspect ratios and the dipoles' orientation. We calculate the species-independent stability diagram and the deformation of the Fermi surface (FS) for polarized molecules, whose electric dipoles are oriented along a preferential direction. Compared to atomic magnetic species, the stability of a molecular electric system turns out to strongly depend on its geometry and the FS deformation significantly increases.
\end{abstract}

\maketitle

The Fermi surface (FS) is one of the fundamental pillars of modern condensed matter physics \cite{R1}. It represents the surface in reciprocal space, which separates occupied from unoccupied fermionic states at zero temperature, and is a direct consequence of the Pauli exclusion principle. For instance, interacting electrons in a normal metal can be described within the Landau Fermi-liquid theory \cite{pom2} as noninteracting fermionic quasi-particles with an effective mass, whose ground state forms such a FS. Due to the isotropy of the Coulomb repulsion between electrons in a uniform space, the FS turns out to be a sphere, whose radius is given by the Fermi momentum. The concept of the FS is crucial for understanding transport processes in metals \cite{R3} and the Cooper pairing in superconductors \cite{R4, Bennemann}. However, in case of complex interactions the FS can get modified. For example, in strongly-correlated electron systems the Fermi-liquid picture breaks down, giving rise to a spontaneous breaking of rotational invariance, which manifests itself in a deformation of the FS \cite{R5}.

Studying Fermi surfaces has now also become accessible within the realm of ultracold quantum gases \cite{BlochDalibardZwerger,Stringari-RMP,PfauRep,Blume, Massignan} due to their high degree of tunability. In Fermi gases consisting of atoms or molecules with a permanent or induced magnetic or electric dipole moment the anisotropic and long-range dipole-dipole interaction (DDI) competes with the large kinetic energy close to the FS \cite{Baranov}. As a consequence, many theoretical papers predicted an anisotropic version of the Landau Fermi-liquid theory \cite{sarma, bar1, bar2}, which involves a deformation of the Fermi sphere \cite{goral, Sogo1, Sogo2, Lima1, Lima2, Blakie}. A recent experiment \cite{Francesca, Veljic2} measured that for a fermionic gas of magnetic dipolar erbium atoms an ellipsoidal deformation of the Fermi sphere occurs, which is of the order of 2\%. This is expected to lead to novel many-body phenomena, in particular in connection with fermionic superfluidity \cite{pairing1,pairing2,pairing3,pairing4, FrancescaPRL, HuiHu}. In a polarized one-component Fermi gas an intriguing interplay between an anisotropic order parameter with odd partial waves and the FS deformation enhances superfluid pairing via modifying the density of states \cite{pairing3}. In contrast to that the more conventional type of Cooper pairing is predicted in a two-component dipolar Fermi gas, where the usual BCS theory together with the deformed FS leads to both spin-singlet even partial wave or spin-triplet odd partial wave Cooper pairs \cite{pairing4}. And it is suggested to obtain and observe a topological $p$-wave superfluid of microwave-dressed polar fermionic molecules in 2D lattices at temperatures of the order of tens of nanokelvins~\cite{N9}.

Since the first experimental realization of a quantum degenerate dipolar Fermi gas of $^{161}$Dy in 2012 \cite{Dy}, several more fermionic species, such as $^{167}$Er \cite{Er} and $^{53}$Cr \cite{Cr}, were successfully cooled down to quantum degeneracy, which enabled studies of the effects of weak to medium-range DDI strength. However, the study of the strongly dipolar regime is still in its infancy, and awaits experimental availability of ultracold heteronuclear polar molecules with large dipole moments. In the last decade, significant efforts to produce chemically stable cold polar molecules \cite{Carr, Gadway} were based on photoassociation or the stimulated Raman adiabatic passage (STIRAP) \cite{Bergmann}. As a result, samples of fermionic $^{40}$K$^{87}$Rb \cite{KRb}, $^{23}$Na$^{40}$K \cite{NaK1, NaK2, NaK3, NaK4}, $^{23}$Na$^{6}$Li \cite{NaLi} and bosonic $^{7}$Li$^{133}$Cs \cite{LiCs1, LiCs2}, $^{87}$Rb$^{133}$Cs \cite{RbCs1, RbCs2} and $^{23}$Na$^{87}$Rb \cite{NaRb} were obtained in deeply bound molecular states. However, the quantum degeneracy was still not reached. Only very recently a quantum degenerate dipolar Fermi gas of $^{40}$K$^{87}$Rb has been realized at JILA \cite{KRb-qd}. This experimental protocol enabled to produce tens of thousands of unpolarized molecules at a temperature as low as 50~nK, which are well described by the Fermi-Dirac distribution. However, the molecules' dipoles can be straight-forwardly polarized in a preferential direction by an external electric field \cite{KRb-qd}, such that the DDI dominates the behavior of the system. This would be a long-awaited significant step forward, which would open up the realm for experimentally investigating strong dipolar Fermi gases.

The stability of quantum degenerate dipolar Fermi gases against mechanical collapse is defined by the positivity of the compressibility due to the Pomeranchuk criterion, which is a special case of the well-known criterion of thermodynamic stability \cite{pom1, pom2}. It was previously considered in cylindrically symmetric harmonic traps \cite{goral, Sogo1, Sogo2, Lima1, Lima2, Blakie}, as well as in homogeneous systems \cite{Bohn}. Here we study the ground state stability of ultracold Fermi gases with tilted dipoles in triaxial harmonic traps and reveal a universal behavior of the critical DDI strength. In particular, we investigate the stability of a polarized $^{40}$K$^{87}$Rb gas in an experimentally realistic parameter regime and calculate critical values of the electric dipole moment and the corresponding FS deformation. Note that, in contrast to non-dipolar systems, the 3-body recombination of dipolar atoms presumably does not play an important role in determining the stability of the system in the parameter range of current experiments \cite{KRb-qd}. Finally, we demonstrate effects of the strong DDI on the time-of-flight expansion and show that a nonballistic expansion theory is essential to understand the dynamical behavior of strongly dipolar Fermi gases.

To achieve this, we use a variational phase-space approach \cite{Falk, Veljic1, Veljic2} for the Wigner function $\nu({\mathbf r},\, {\mathbf k})=\int d^3 r'\, e^{-i {\mathbf k}\cdot {\mathbf r}'}\, \rho\left({\mathbf r}+\frac{1}{2}{\mathbf r}',\, {\mathbf r}-\frac{1}{2}{\mathbf r}'\right)$, which relies on the Hartree-Fock mean-field approximation. Here $\rho({\mathbf r},\, {\mathbf r}')=\langle\hat\Psi({\mathbf r})\hat\Psi^\dagger({\mathbf r}')\rangle$ represents the one-body density matrix. Note that the second-order terms in the DDI in the theory beyond Hartree-Fock \cite{Kopietz1,Kopietz2} yield only a small correction even for polar molecules, although the geometry may have an impact (see Supplemental Material \cite{SuppMat} for more details). Furthermore, this beyond-mean-field correction turns out to destabilize the system \cite{Kopietz1, Kopietz2}, so our results on the stability represent proper upper boundaries. This is in stark contrast to bosonic systems, where the quantum fluctuations have turned out to stabilize the system and lead, for instance, to the formation of quantum droplets \cite{al1,droplet1, droplet2, fl1, droplet3, droplet4} and supersolids \cite{droplet5, modugno, pfausupersolid, supsol1, supsol2, supsol3} in Bose-Einstein condensates of dysprosium and erbium in Stuttgart, Innsbruck, and Pisa experiments.

\begin{figure}[!t]
\centering
\includegraphics[width=6.5cm]{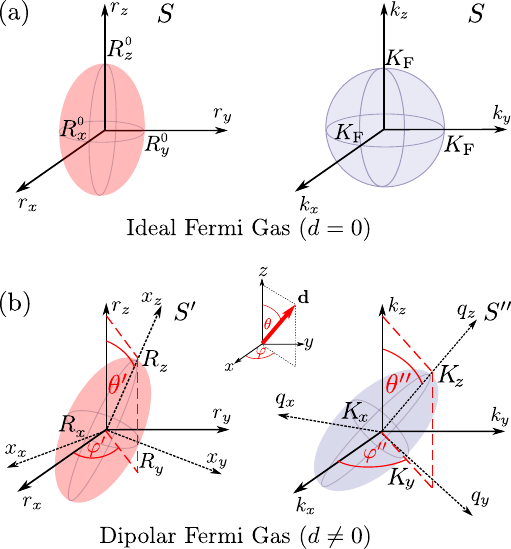}
\caption{(a) Shape of the molecular cloud and the FS of ideal Fermi gas. (b) Schematic illustration of the ansatz for the shape of the molecular cloud and the FS of dipolar Fermi gas. The inset shows orientation of dipoles.}
\label{fig:fig1}
\end{figure}

We consider the dipolar Fermi system to be at zero temperature. This is justified as the temperature in the experiments  \cite{KRb-qd} is about $T/T_F\approx 0.3$ and as thermal corrections to the total energy are proportional to $(T/T_F)^2$ \cite{Howe}. Thus, we assume a Heaviside-shaped Wigner function in the ground state and obtain the total energy of $N$ identical fermions of mass $M$ in terms of the Thomas-Fermi (TF) momenta $K_i$ and radii $R_i$,
\begin{flalign*}
&\hspace*{-3mm}E_{\rm tot}=\frac{N}{8}\left(\sum_{i}\frac{\hbar^2K_i^2}{2M}+\sum_{i,j}\frac{M \omega_i^2  \mathbb{R}_{ij}'^2 R_j^2}{2}\right)-\frac{6N^2 c_0}{R_x R_y R_z}\\
&\hspace*{-3mm}\times \left[F_A\left(\frac{R_x}{R_z},\frac{R_y}{R_z},\theta,\varphi, \theta',\varphi'\right) -F_A\left( \frac{K_z}{K_x},\frac{K_z}{K_y}, \theta,\varphi, \theta'',\varphi'' \right) \right] .
\end{flalign*}
Here the angles ($\theta,\varphi$) determine the dipoles' orientation, ($\theta',\varphi'$) describe the orientation of the molecular cloud, and ($\theta'',\varphi''$) determine the FS orientation, as illustrated in Fig.~\ref{fig:fig1}. We stress that the molecular cloud orientation does not have to coincide with the trap orientation, due to the DDI effects, and $\mathbb{R}'$ stands for the corresponding rotation matrix, while $\omega_i$ denote the trap frequencies and $F_{\rm A}$ is a generalized anisotropy function (see Supplemental Material \cite{SuppMat} for further details). Note that with this our theory not only allows a quantitative analysis of current experimental data. In addition, it also provides a framework for disentangling reliably the rigid rotation of the Fermi ellipsoid, which occurs for the weak DDI in magnetic dipolar atoms in Innsbruck  \cite{Francesca, Veljic2}, from its deformation in case of the strong DDI in electric dipolar molecules investigated at JILA \cite{KRb-qd}.

\begin{figure*}[!t]
\centering
\includegraphics[height=4cm]{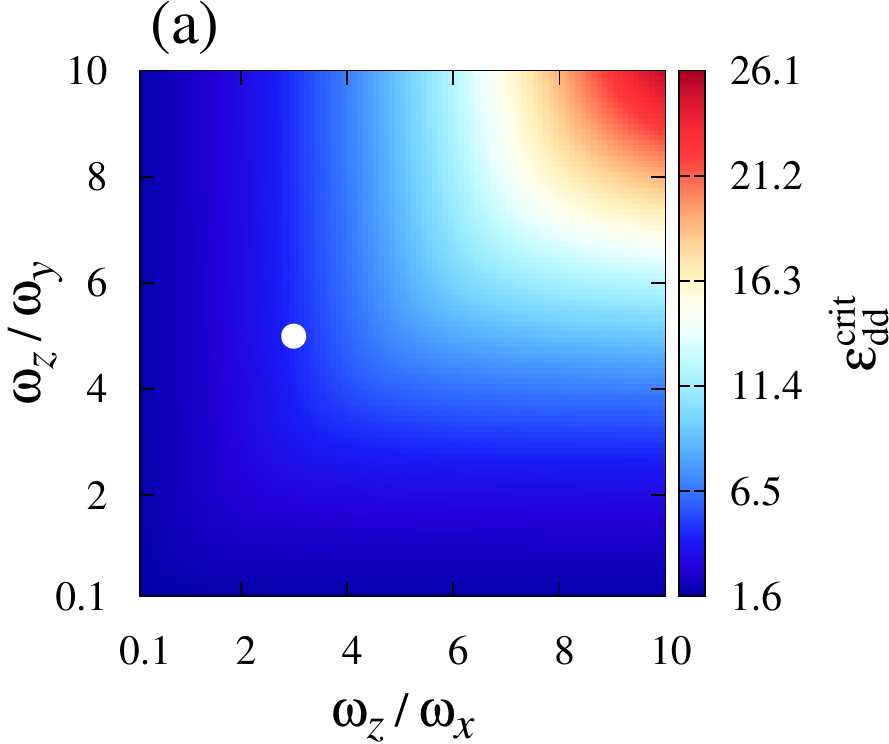}\hspace*{2mm}
\includegraphics[height=4cm]{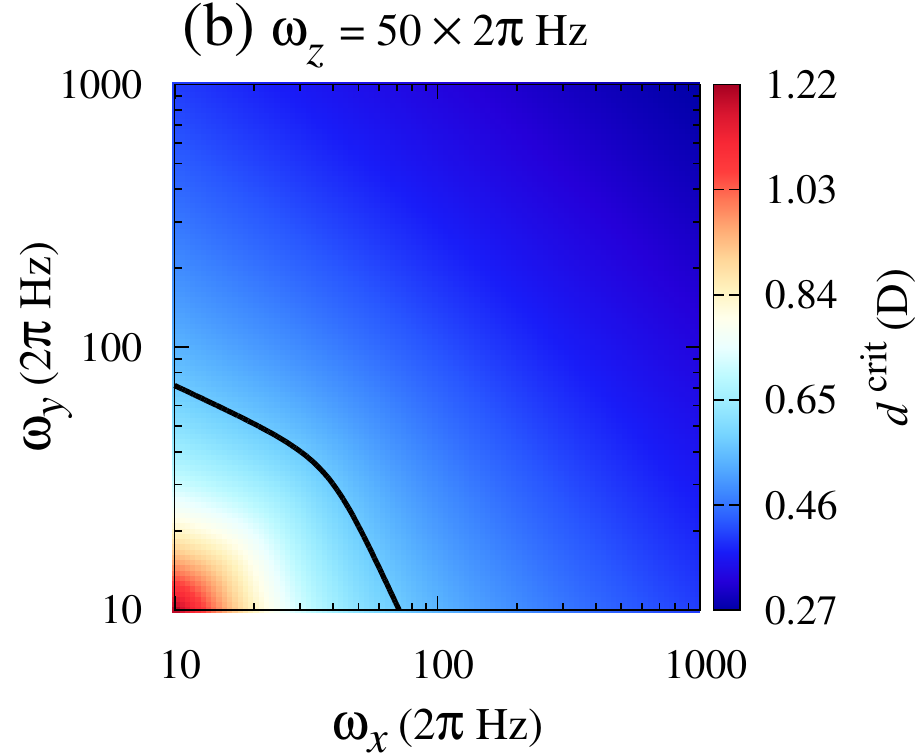}\hspace*{2mm}
\includegraphics[height=4cm]{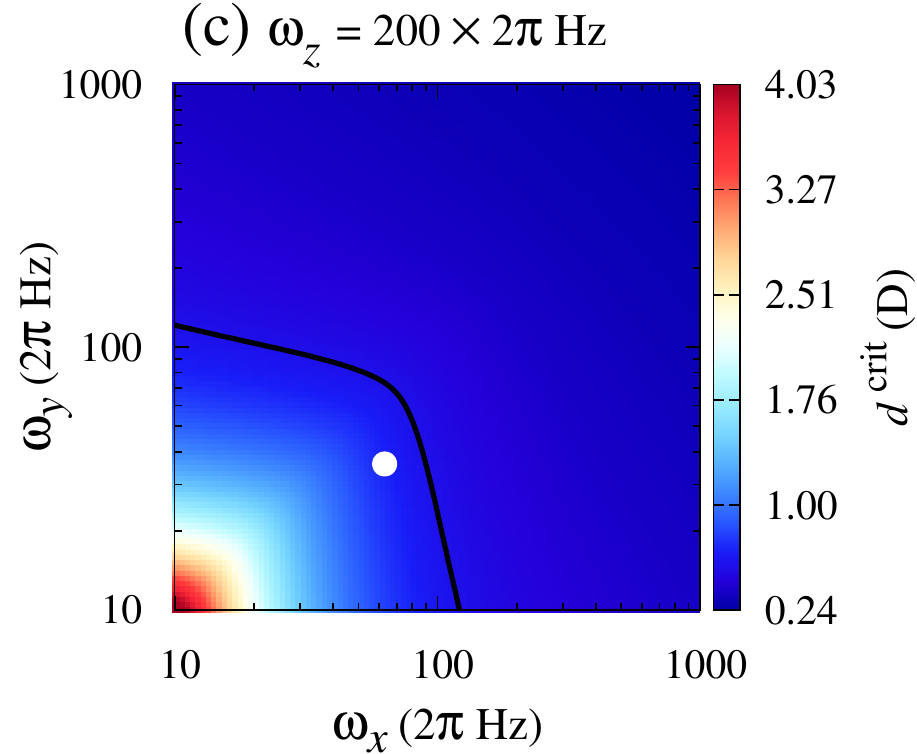}
\caption{(a) A universal stability diagram for harmonically trapped ultracold dipolar Fermi gases at quantum degeneracy: critical value of the relative dipole-dipole interaction strength $\varepsilon_\mathrm{dd}^\mathrm{crit}$ for $\theta=\varphi=0$. The system has a stable ground state for $\varepsilon_\mathrm{dd}\leq\varepsilon_\mathrm{dd}^\mathrm{crit}$. (b), (c) Critical value of the electric dipole moment $d^\mathrm{crit}$ for a stable ground state of $N=3\times 10^4$ ultracold molecules of $^{40}$K$^{87}$Rb for $\theta=\varphi=0$: (b) $\omega_z=2\pi\times 50$~Hz; (c) $\omega_z=2\pi\times 200$~Hz. White dots in (a), (c) correspond to the parameters of experiment \cite{KRb-qd}, while black lines in (b), (c) correspond to the permanent dipole moment $d=0.574$~D of $^{40}$K$^{87}$Rb molecules.}
\label{fig:fig2}
\end{figure*}

By extremizing the above energy with respect to the variational parameters  ($R_i, K_i, \theta',\varphi', \theta'', \varphi''$), we obtain the corresponding equations for the ground state, which can be rewritten in a dimensionless, species-independent form \cite{SuppMat} such that, for a given orientation of the dipoles, they only depend on three parameters: the two trap aspect ratios $\omega_z/\omega_x$ and $\omega_z/\omega_y$, and the relative DDI strength
\begin{equation*}
\varepsilon_\mathrm{dd}=\frac{d^2}{4\pi\varepsilon_0}\sqrt{\frac{M^3}{\hbar^5}}(\omega_x \omega_y \omega_z N)^{1/6}\, ,
\end{equation*}
where $d$ denotes the electric dipole moment. This remarkable result reveals a universality governing the ground-state properties of dipolar Fermi gases. Furthermore, it allows us to determine the stability diagram of the system, shown in Fig.~\ref{fig:fig2}(a) for the case $\theta=\varphi=0$, in terms of the maximal DDI strength $\varepsilon_\mathrm{dd}^\mathrm{crit}$ for which the ground state exists. We see that large aspect ratios significantly increase the critical DDI strength, thus stabilizing the system in a much broader parameter range. As an immediate consequence we read off from Fig.~\ref{fig:fig2}(a), for instance, that a dipolar Fermi gas can be stabilized against mechanical collapse, which arises in 3D for a sufficiently strong interaction, by confining the polar molecules to 2D, i.e., to a monolayer. Note that using the tilting angle of the dipole orientation relative to the monolayer and the DDI as control parameters, one can find apart from the normal Fermi liquid and the collapse also a superfluid phase and a density-wave phase \cite{bar1, bar2}.

We also note that $\varepsilon_\mathrm{dd}^\mathrm{crit}$ turns out to be a symmetric function of its arguments $\omega_z/\omega_x$ and $\omega_z/\omega_y$ \cite{SuppMat}.
If we consider the experimentally available species $^{40}$K$^{87}$Rb, the stability diagram from Fig.~\ref{fig:fig2}(a) can be used to obtain a species-specific stability diagram for a particular value of one of the trap frequencies, as shown in Figs.~\ref{fig:fig2}(b) and \ref{fig:fig2}(c). Here we see how the critical value of the dipole moment $d^\mathrm{crit}$ depends on $\omega_x$ and $\omega_y$ for a fixed value of $\omega_z$. If we take into account that the permanent electric dipole moment of  $^{40}$K$^{87}$Rb has the value $d=0.574$~D, denoted by black lines in Figs.~\ref{fig:fig2}(b) and \ref{fig:fig2}(c), we read off that for $\omega_z=2\pi\times 50$~Hz the instability can kick in already for frequencies $\omega_x$, $\omega_y$ of that order or larger. In the experiment of Ref.~\cite{KRb-qd} the frequencies used are $(\omega_x,\, \omega_y,\, \omega_z)=2\pi\times (63,\, 36,\, 200)$~Hz, and Fig.~\ref{fig:fig2}(c) reveals that the system may easily become unstable for slightly larger frequencies if the dipoles would be polarized along $z$ axis.

The most striking effect that can be demonstrated in the strong DDI regime is the FS deformation $\Delta=K_z/K_x-1$, defined in terms of the TF momenta aspect ratio. It was experimentally observed for the first time for magnetic dipolar $^{167}$Er atoms \cite{Francesca}, where $\Delta$ is of the order of a few percent. This effect is much larger in gases of polar molecules, as can be seen for $^{40}$K$^{87}$Rb in Fig.~\ref{fig:fig3}(a), obtained by solving the equations presented in the Supplemental Material \cite{SuppMat}. Here it is assumed that the electric dipole moment is tuned down to $d=0.22$~D, such that it is below the minimal value of $d^\mathrm{crit}=0.24$~D obtained in Fig.~\ref{fig:fig2}(c). For realistic values of the trapping frequencies we obtain that $\Delta$ varies between 5\% and 30\%. 

Furthermore, the theory presented here makes it possible to calculate the stability properties for experimentally relevant dipolar Fermi systems, where even relatively small changes in the dipolar moment strength can significantly affect the system's stability. This is demonstrated in Fig.~\ref{fig:fig3}(b), where for a slightly larger value of $d=0.26$~D we read off that the FS deformation becomes significantly larger than in Fig.~\ref{fig:fig3}(a), namely up to 45\%, and that an unstable region appears, which does not support a stable ground state of the system.

\begin{figure}[!b]
\centering
\includegraphics[height=4.2cm]{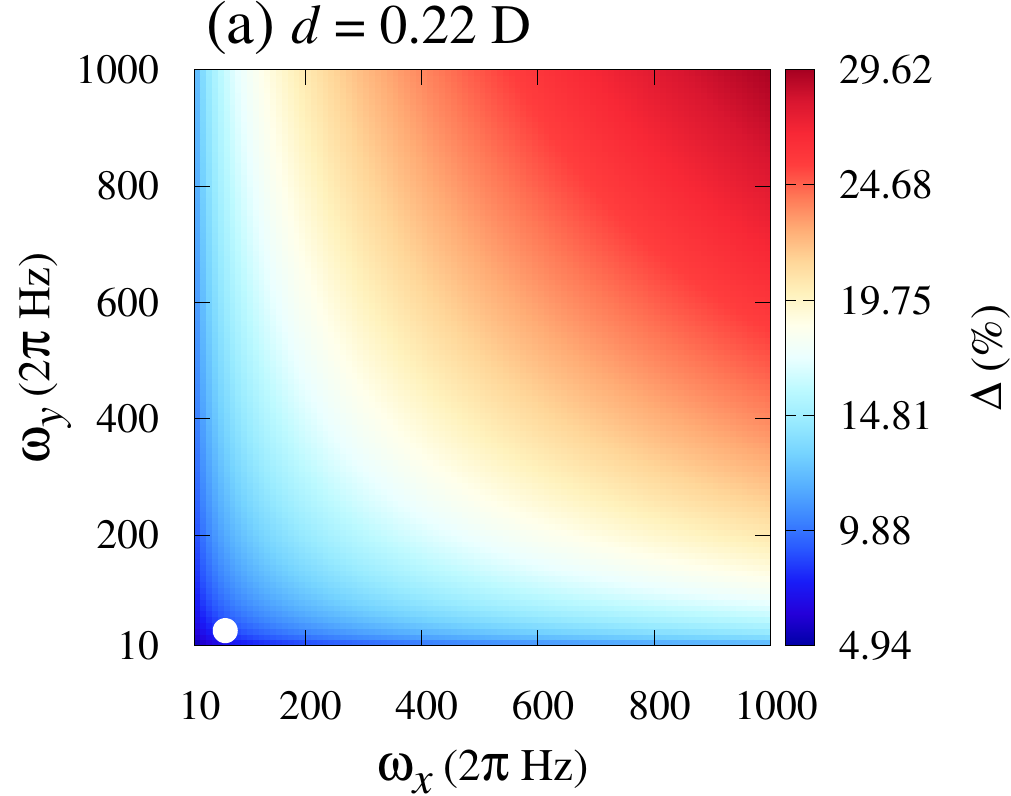}\vspace*{2mm}
\includegraphics[height=4.2cm]{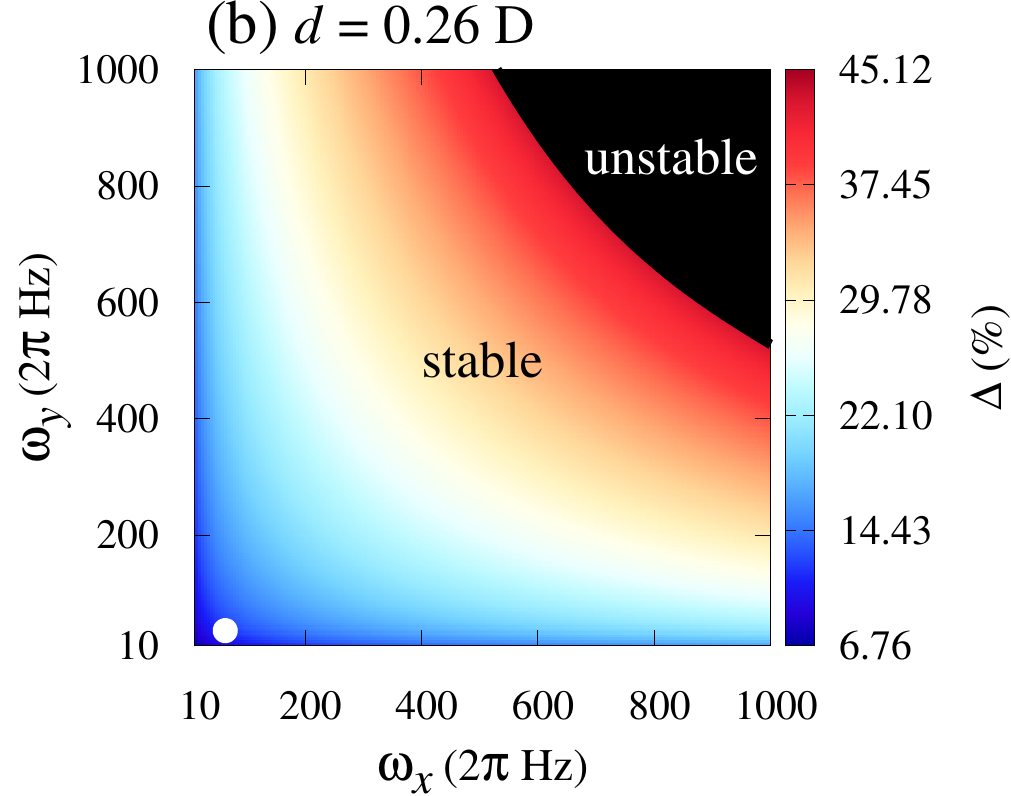}
\caption{FS deformation $\Delta$ as a function of $\omega_x$ and $\omega_y$ for a $^{40}$K$^{87}$Rb system with $N=3\times 10^4$, $\omega_z=2\pi\times 200$~Hz, $\theta=\varphi=0$, and electric dipole moments: (a) $d=0.22$~D; (b) $d=0.26$~D. White dots correspond to the parameters of Ref.~\cite{KRb-qd}. Black region in (b) does not yield stable solutions.}
\label{fig:fig3}
\end{figure}

\begin{figure*}[!t]
\centering
\includegraphics[height=4.2cm]{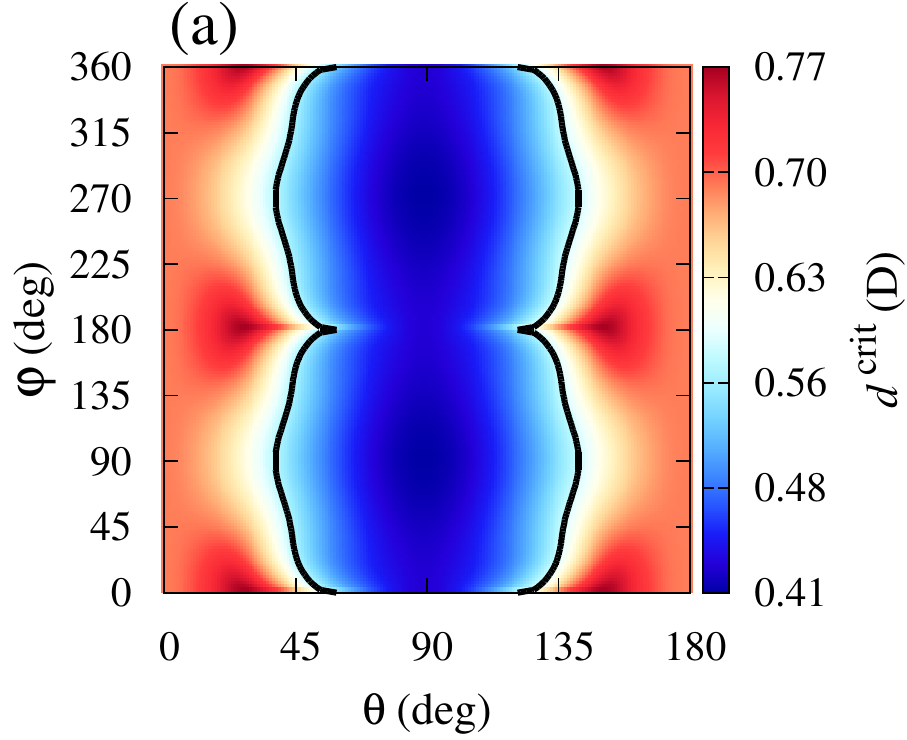}\hspace*{2mm}
\includegraphics[height=4.2cm]{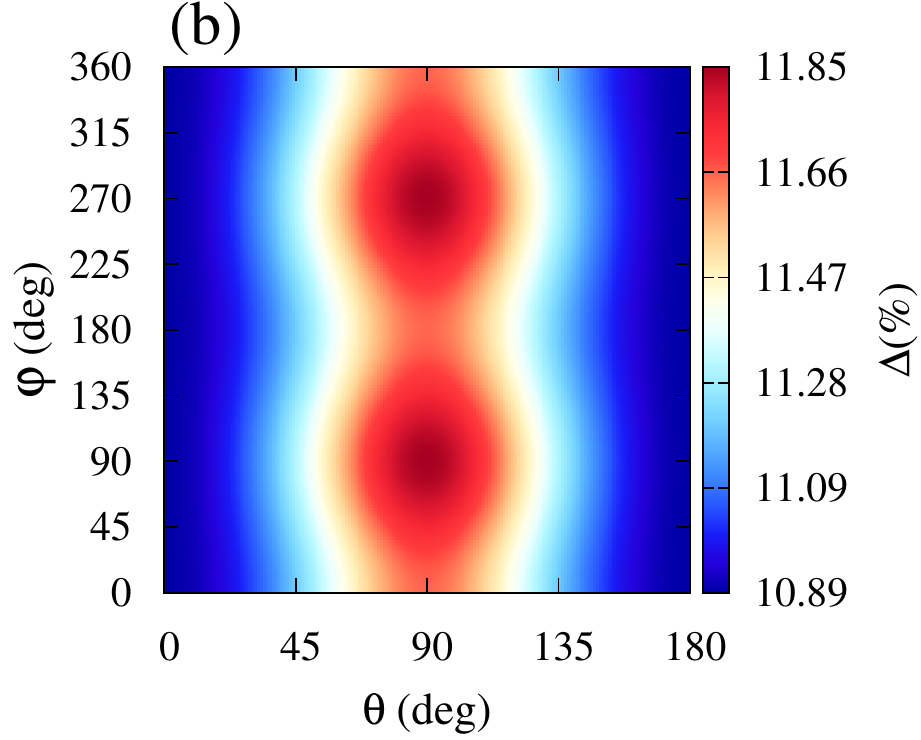}\hspace*{2mm}
\includegraphics[height=4.2cm]{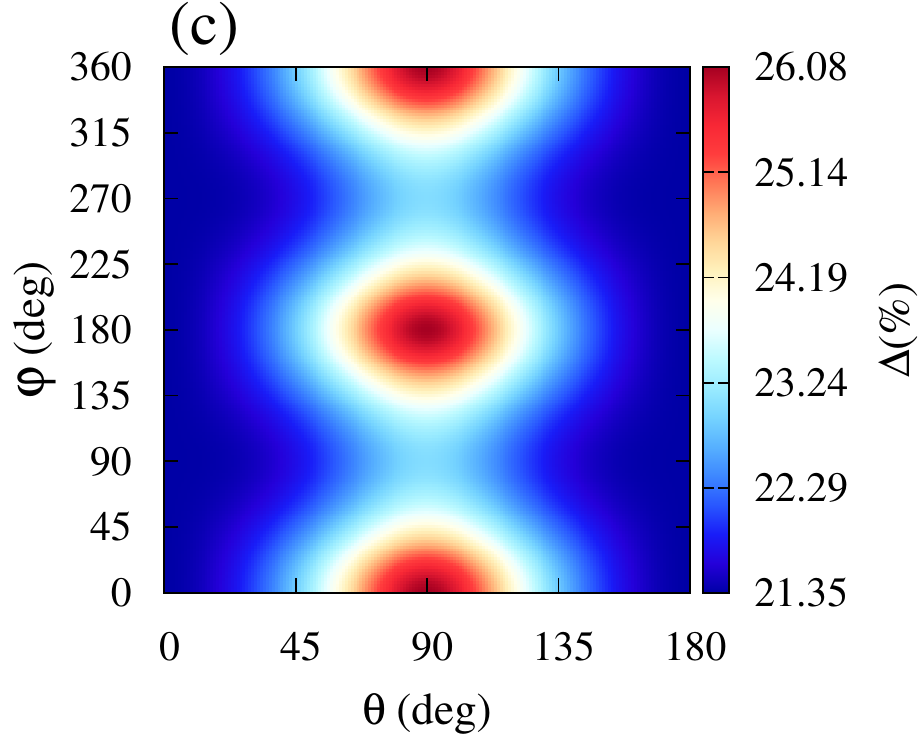}
\caption{(a) Angular stability diagram of $^{40}$K$^{87}$Rb: $d^\mathrm{crit}$ as a function of the dipoles' orientation. Black line corresponds to the permanent electric dipole moment $d=0.574$~D of $^{40}$K$^{87}$Rb, and the trap frequencies are as in Ref.~\cite{KRb-qd}, with $N=3\times 10^4$ molecules. (b), (c) Angular dependence of $\Delta$ for a fixed value $d=0.25$~D and the trap frequencies: (b) as in Ref.~\cite{KRb-qd}; (c) $(\omega_x,\, \omega_y,\, \omega_z)=2\pi\times (50,\, 500,\, 900)$~Hz.}
\label{fig:fig4}
\end{figure*}

Previously it was always assumed \cite{goral, Sogo1, Sogo2, Lima1, Lima2, Blakie, Falk, Veljic1, Veljic2} that the cloud shape in real space follows the trap orientation, which is a reasonable approximation for a weak DDI and an elongated trap. However, here we provide a theory capable to describe dipolar Fermi systems in a general trap geometry for all DDI strengths. Therefore, we leave the orientation angles ($\theta',\varphi'$) of the TF ellipsoid in real space as free variational parameters, together with the TF radii $R_i$, as illustrated in Fig.~\ref{fig:fig1}(b). It was previously experimentally verified and theoretically always assumed \cite{Francesca,Veljic2} that the FS stretches into an ellipsoid along the orientation of the dipoles, as can be expected due to symmetry reasons. Here we use a more general ansatz, where the FS orientation angles ($\theta'',\varphi''$) are also taken as free variational parameters, together with the TF momenta $K_i$, Fig.~\ref{fig:fig1}(b). However, we show here \cite{SuppMat} that the principle of minimizing the energy leads to the solution $\theta''=\theta$, $\varphi''=\varphi$ \cite{SuppMat}, i.e., that our theory confirms the notion that the FS follows the dipoles' orientation and properly captures the physical behavior of the system. This result also allows us to reduce the number of variational parameters to eight, ($R_i$, $K_i$, $\theta'$, $\varphi'$), as well as the number of equations \cite{SuppMat}.

The cloud orientation obtained within our theory strongly depends on both the DDI strength and the elongation of the trap. In the special case of a spherical trap the cloud is elongated along the dipoles' direction, as the FS, but in a general case the cloud orientation can only be determined numerically. Figure~\ref{fig:fig4}(a) shows the angular stability diagram for a $^{40}$K$^{87}$Rb system in terms of the critical dipole moment $d^\mathrm{crit}$, where all variational parameters ($R_i$, $K_i$, $\theta'$, $\varphi'$) are numerically calculated for each configuration. If one would assume that the molecular cloud follows the trap shape, i.e., $\theta'=\varphi'=0$, the obtained values of $d^\mathrm{crit}$ would be significantly different from those calculated in Fig.~\ref{fig:fig4}(a) (for a comparison, see Supplemental Material \cite{SuppMat}). This demonstrates that the theory developed here is important for an accurate qualitative and quantitative description of dipolar Fermi systems with moderate to strong DDI. Figures~\ref{fig:fig4}(b) and \ref{fig:fig4}(c) illustrate that the trap geometry also strongly affects the system's behavior, and that the FS deformation and its angular distribution can be tuned by changing the trap frequencies. Not only the range of the FS deformation values can be increased or decreased this way, but also its minima and maxima can be freely modified. The observed strong angular dependence of the FS deformation has an important consequence, namely, that the FS does not only follow the dipoles' orientation, but its shape gets modified as well. This is a qualitatively different behavior compared to atomic magnetic species, where the angular dependence of the FS deformation is quite weak \cite{Veljic2}, thus the FS just rigidly follows the dipoles' orientation.

Effects of the DDI and their interplay with the geometry also quite strongly influence the dynamics of the system, which is of particular importance for interpreting experimental time-of-flight (TOF) imaging data \cite{Veljic1, Veljic2}. For polar molecules with a strong DDI the difference between the usually assumed ballistic and the actual nonballistic expansion can be huge, which we show here. The TOF expansion imaging is commonly used for experimental measurements of the properties of ultracold Fermi gases, and the deformation of the cloud shape is described in terms of the cloud aspect ratio $A_R(t)$, which is defined by the ratio of the average sizes of the cloud in vertical $\sqrt{\langle r^2_\mathrm{v}\rangle}$ and horizontal $\sqrt{\langle r^2_\mathrm{h}\rangle}$ direction in the imaging plane. Since the imaging axis in the experiment of Ref.~\cite{KRb-qd} lies in the $xy$ plane and forms an angle $\gamma=22.5^\circ$ with respect to the $x$ axis, according to Ref.~\cite{Veljic1,Veljic2} the aspect ratio is given by
\begin{equation*}
\label{eq:AR}
A_R(t)=\sqrt{\frac{\langle r^2_\mathrm{v}\rangle}{\langle r^2_\mathrm{h}\rangle}}=\frac{R_z b_z(t)}{\sqrt{R_x^2b^2_x(t)\sin^2\gamma+R_y^2b^2_y(t)\cos^2\gamma}}\, ,
\end{equation*}
where the scaling parameters $b_i(t)$ represent variations from the global equilibrium values of the TF radii and momenta \cite{Stringari-ansatz}. A detailed derivation of equations of motion for the scaling parameters based on the quantum Boltzmann equation within the self-consistent relaxation-time approach for $\theta=\varphi=0$ is given in Ref.~\cite{Veljic1}. Here we numerically solve these equations for the general triaxial trap geometry and the parameters corresponding to the polar molecules of Ref.~\cite{KRb-qd}.

\begin{figure}[!t]
\centering
\includegraphics[height=4.5cm]{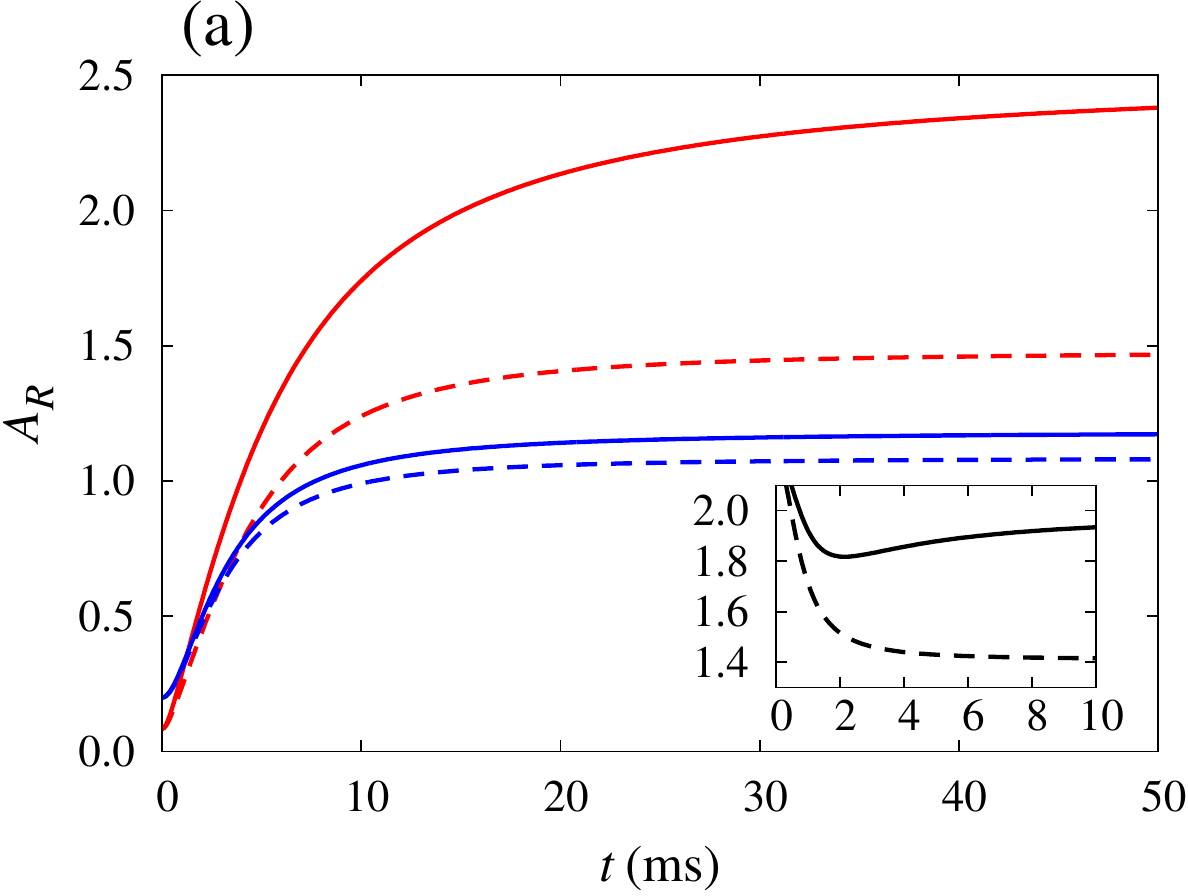}\vspace*{2mm}
\includegraphics[height=4.5cm]{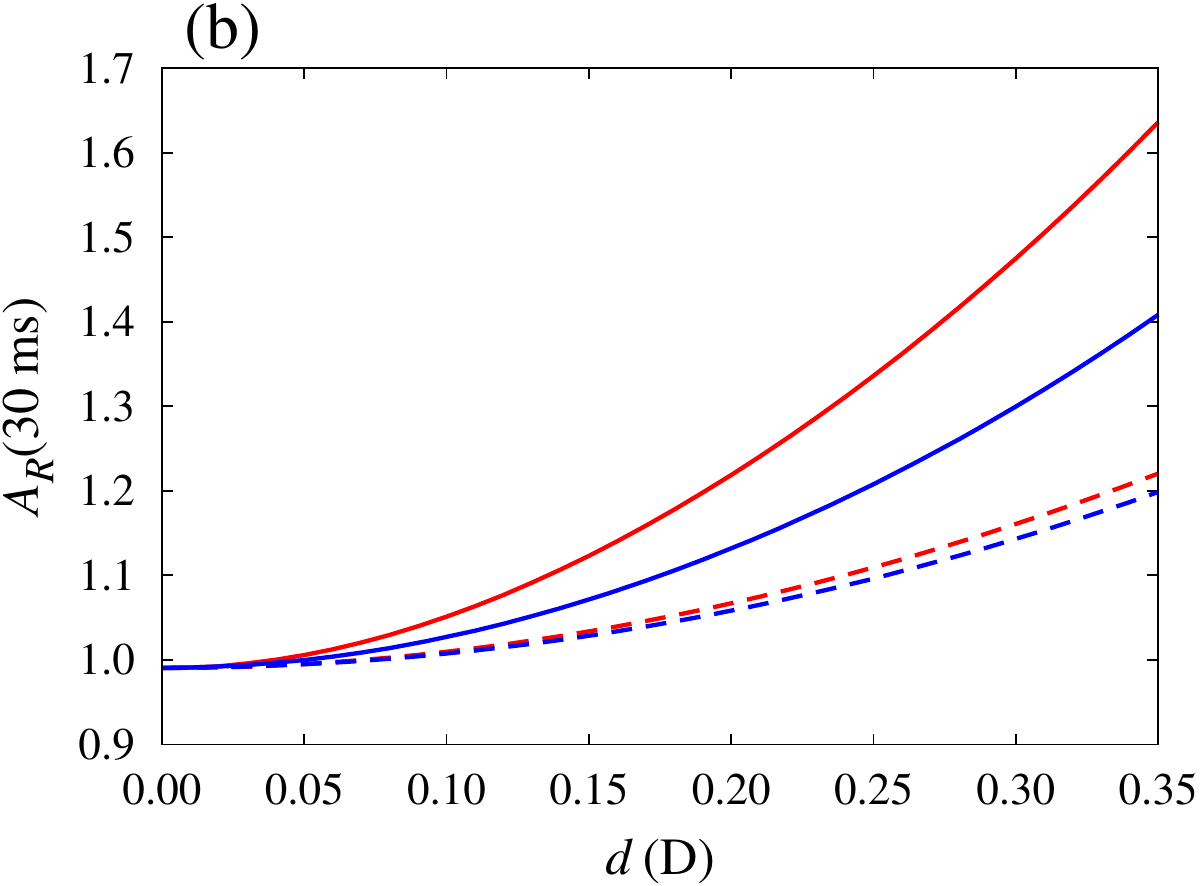}
\caption{(a) Real-space aspect ratio $A_R$ of the $^{40}$K$^{87}$Rb molecular cloud as a function of time $t$ during the TOF expansion from the ground state, after the trap is switched off. Top (red) solid and dashed lines are obtained for $d=0.5$~D and frequencies $2\pi\times (63,\, 36,\, 500)$~Hz, bottom (blue) solid and dashed lines for $d=0.22$~D and $2\pi\times (63,\, 36,\, 200)$~Hz, and inset for $d=0.35$~D and $2\pi\times (250,\, 150,\, 100)$~Hz. (b) $A_R$(30~ms) after the TOF expansion for $t=30$~ms as a function of $d$. Trap frequencies corresponding to all line types are the same as in (a). In both plots solid lines correspond to a nonballistic expansion, where the DDI is taken into account, while the dashed lines represent calculated results for a free (ballistic) expansion, $N=3\times 10^4$, $\theta=\varphi=0$. $A_R$ is calculated using the imaging angle $22.5^\circ$ of Ref.~\cite{KRb-qd}, in the geometry of Ref.~\cite{Veljic2}.}
\label{fig:fig5}
\end{figure}

It was previously shown that, even for magnetic atomic species such as erbium, the DDI effects could be experimentally observed in the TOF dynamics, and that a nonballistic expansion has to be used in order to properly describe the system's behavior \cite{Veljic1, Veljic2}. For polar molecules with a strong DDI we expect that nonballistic effects are more pronounced, as can be read off from Fig.~\ref{fig:fig5}. Even more significant are large variations of nonballistic effects, which can be as small as 8\% or as large as 60\% for quite similar configurations, as is illustrated for the two examples in Fig.~\ref{fig:fig5}(a). Although the trap geometry plays a role here, Fig.~\ref{fig:fig5}(b) reveals that the ballistic behavior is roughly the same, as is expected based on the system parameters, while the DDI strength gives a major contribution. Furthermore, the inset in Fig.~\ref{fig:fig5}(a) shows that even the qualitative behavior of the system can be incorrectly predicted (monotonous vs.~nonmonotonous behavior) when nonballistic effects are neglected. This demonstrates that the DDI has to be taken into account even during the TOF expansion, and that the interpretation of experimental data are hugely affected by the model used. Therefore, the general theory presented here also enables an accurate modeling of the dynamics of strongly interacting dipolar Fermi systems.

In conclusion, we have presented a general mean-field theory for the ground state of polarized, harmonically trapped dipolar Fermi gases at zero-temperature, with an arbitrary orientation of the dipoles. We have derived a universal, species-independent set of equations for the ground state and investigated the stability of systems of polar molecules. We have shown that the molecular cloud shape and the FS deformation strongly depend on the dipoles' orientation. Our results are important for the study of the interplay between the FS deformation and superfluid pairing \cite{pairing1,pairing2,pairing3,pairing4}, in particular to address the open question of how the anisotropic order parameter of the emergent superfluidity and its critical temperature are tunable by both the trap geometry and the dipoles' orientation. The presented theory paves the way towards new methods for quantum engineering of properties of dipolar Fermi gases that depend on the FS shape, such as the emergence of superfluidity. In the outlook we mention that possible fermionic analogues of the already observed bosonic quantum droplets may exist. Although the leading beyond-mean-field correction seems to destabilize fermionic systems, further studies might reveal an alternative stabilization mechanism.

\textit{Acknowledgements.}
We acknowledge inspiring discussions with J.~P. Covey, A.~R.~P.~Lima, L.~De Marco, K.~Matsuda, W.~Tobias, G.~Valtolina, I.~Vasi\'{c}, and J.~Ye.
This work was supported in part by the Ministry of Education, Science, and Technological Development of the Republic of Serbia under projects ON171017 and QDDB, by the German Academic and Exchange Service (DAAD) under project QDDB, and by the Deutsche Forschungsgemeinschaft (DFG, German Research Foundation) via the Collaborative Research Centers SFB/TR185 (project No.~277625399) and SFB/TR49, as well as the Research Unit FOR 2247 (project No.~PE 530/6-1). Numerical simulations were run on the PARADOX supercomputing facility at the Scientific Computing Laboratory of the Institute of Physics Belgrade.

\bibliography{KRb-references}{}

\onecolumngrid
\setcounter{figure}{0}
\setcounter{equation}{0}
\renewcommand{\theequation}{S\arabic{equation}}
\renewcommand{\thefigure}{S\arabic{figure}}
\appendix

\vspace*{5mm}
\begin{center}
\large\bf
Supplemental material:\\
Stability of quantum degenerate Fermi gases of tilted polar molecules
\end{center}

\section{Global equilibrium}

We consider an ultracold quantum degenerate dipolar Fermi gas at zero temperature in global equilibrium. The system consists of $N$ identical spin-polarized single-component fermions of mass $M$ with an electric dipole moment ${\bf d}$. The system is confined into a three-dimensional harmonic trap with the frequencies $\omega_i$, whose axes coincide with the axes of the laboratory coordinate system $S$, as it is depicted in Fig.~\ref{fig:fig1}. The dipole moments are aligned in the direction defined by spherical angles $(\theta, \varphi)$, as shown in Fig.~\ref{fig:fig1}(b). For the ideal Fermi gas, the molecular cloud shape is determined purely by the trap, while the FS is a sphere, Fig.~\ref{fig:fig1}(a). In contrast to this, when the DDI is present, the molecular cloud shape will depend on both the trap geometry and the orientation of the dipoles. Here we assume that it has an ellipsoidal shape, oriented along the direction defined by the angles $(\theta',\varphi')$, as depicted in Fig.~\ref{fig:fig1}(b). Similarly, the FS is stretched into an ellipsoid whose longitudinal axis is expected to coincide with the dipoles' orientation, but we do not assume it from the beginning and will instead show that this can be obtained within the presented theory. Therefore, as depicted in Fig.~\ref{fig:fig1}(b), the FS ellipsoid is oriented in the direction defined by the angles $(\theta'',\varphi'')$, which are free parameters.

We use the following variational ansatz for the Wigner distribution function:
\begin{equation}
 \nu^{}({\bf r},{\bf k})=\Theta\left(1-\sum_{i,j} r_i\mathbb{A}_{ij}r_j-\sum_{i,j} k_i\mathbb{B}_{ij}k_j\right)\, ,
\label{eq:ansatzwigner}
\end{equation}
where $\Theta$ represents the Heaviside step function, while ${\mathbb{A}}_{ij}$ and $\mathbb{B}_{ij}$ are matrix elements that account for the geometry of the system and determine the shape of the cloud in real space and of the FS in momentum space, as illustrated in Fig.~\ref{fig:fig1}(b). With this ansatz we obtain the total energy of the system in the Hartree-Fock approximation:
\begin{align}
E_{\rm tot}=\frac{N}{8}\left(\sum_i\frac{\hbar^2K_i^2}{2M}+\sum_{i,j}\frac{M \omega_i^2  \mathbb{R}_{ij}'^2 R_j^2}{2}\right)-\frac{6N^2 c_0}{R_xR_yR_z}\left[F_A\left( \frac{R_x}{R_z},\frac{R_y}{R_z},\theta,\varphi, \theta', \varphi' \right)-F_A\left( \frac{K_z}{K_x},\frac{K_z}{K_y},\theta,\varphi, \theta'', \varphi'' \right)\right]\, , \label{Etot}
\end{align}
where $R_i$ and $K_i$ are the TF radii and momenta, respectively, $c_0=2^{10}d^2/(\varepsilon_0 \cdot3^4\cdot 5\cdot 7\cdot \pi^3)$ is a constant related to the DDI strength, and ${\mathbb{R}}'_{ij}$ are matrix elements of the rotation matrix $\mathbb{R}'=\mathbb{R}(\theta', \varphi')$, with
\begin{equation}
\mathbb{R}(\alpha,\beta)=\left( \begin{array}{ccc}
\cos\alpha\cos\beta & -\sin\beta & \sin\alpha\cos\beta \\
\cos\alpha\sin\varphi & \cos\beta & \sin\alpha\sin\beta \\
-\sin\alpha & 0 & \cos\alpha \end{array} \right)\, .
\label{eq:mR}
\end{equation}
The features of the DDI are embodied into the generalized anisotropy function $F_A(x,y,\theta,\varphi, \tilde{\theta},\tilde{\phi})$,  which includes the dependence on the general orientation of the dipoles $\bf d$ and the corresponding TF ellipsoid:
\begin{equation}
 F_A(x,y,\theta,\varphi,\tilde{\theta},\tilde{\phi})=\left({\sum_i\mathbb{R}_{iz}  \tilde{\mathbb{R}}_{ix}}\right)^2f\left(\frac{y}{x},\frac{1}{x}\right)+\left({\sum_i\mathbb{R}_{iz}  \tilde{\mathbb{R}}_{iy}}\right)^2f\left(\frac{x}{y},\frac{1}{y}\right)+\left({\sum_i\mathbb{R}_{iz}  \tilde{\mathbb{R}}_{iz}}\right)^2 f(x,y)\, ,
\label{gen_aniso_func}
\end{equation}
where $\mathbb{R}_{ij}$ and $\tilde{\mathbb{R}}_{ij}$ are matrix elements of the rotation matrix $\mathbb{R}=\mathbb{R}(\theta, \varphi)$ and $\tilde{\mathbb{R}}=\mathbb{R}(\tilde{\theta},\tilde{\phi})$, respectively. In the above definition, $f(x,y)$ stands for the well-known anisotropy function \cite{pfau_aniso_func} 
\begin{equation}
 f\left(x,y\right)=1-3xy\frac{E(\phi,\kappa)-F(\phi,\kappa)}{(1-y^2)\sqrt{1-x^2}}\, ,
\label{aniso_func}
\end{equation}
where $\phi=\arccos x$ and $\kappa^2=(1-y^2)/(1-x^2)$, while $F(\phi,k)$ and $E(\phi,k)$ are the elliptic integrals of the first and of the second kind, respectively. Note that in the two relevant limiting cases the function $F_A$ satisfies
\begin{equation}
F_A(x,y,0,0,0,0)=f(x,y)\, ,\quad F_A(x,y,\alpha,\beta,\alpha,\beta)=f(x,y)\, .
\end{equation}
Due to these identities and the fact that $f(x,y)$ is a symmetric function, the obtained distributions of $\varepsilon_\mathrm{dd}^\mathrm{crit}$ and $d^\mathrm{crit}$ in Fig.~\ref{fig:fig2}, as well as the distributions of $\Delta$ in Fig.~\ref{fig:fig3} are symmetric with respect to their arguments for $\theta=\varphi=0$. Note that the above definition of the generalized anisotropy function enables symmetric treatment of both the Hartree and the Fock term in the expression for the total energy (\ref{Etot}).

The dipolar Fermi system is determined by the 10 variational parameters ($R_i, K_i, \theta',\varphi', \theta'', \varphi''$), which are obtained by minimizing the total energy (\ref{Etot}), under the constraint that the total number of particles is equal to $N$. This leads to the following set of algebraic equations:
\begin{eqnarray} 
N-\frac{1}{48}R_xR_yR_zK_xK_y K_z&=0\, ,\label{VSTN} \\
K_x-K_y&=0\, , \label{KxKy}  \\
\frac{2\hbar^2K_x^2}{3}-\frac{\hbar^2K_y^2}{3}-\frac{\hbar^2K_z^2}{3}+\frac{48MNc_0}{R_x R_y R_z} \frac{K_z}{K_x}\partial_{K_x}F_A\left(\frac{K_z}{K_x},\frac{K_z}{K_y},\theta,\varphi,\theta'',\varphi'' \right)&=0\, ,
\label{cylindersymmglobalmoment}\\
 \hspace*{-2cm}
\sum_i\omega_i^2  \mathbb{R}_{ix}'^2 R_x^2-\sum_i\frac{\hbar^2{K_i}^2}{3M^2}-\frac{48Nc_0}{MR_x R_y R_z} F_A\left( \frac{K_z}{K_x},\frac{K_z}{K_y}, \theta,\varphi,\theta'',\varphi''\right) \nonumber \\
+\frac{48Nc_0}{MR_x R_y R_z} \left[F_A\left(\frac{R_x}{R_z},\frac{R_y}{R_z},\theta,\varphi,\theta',\varphi' \right)-R_x\partial_{R_x}F_A\left(\frac{R_x}{R_z},\frac{R_y}{R_z},\theta,\varphi,\theta',\varphi' \right)\right]
&=0  \, , \label{VSTRx}\\ 
\hspace*{-2cm}
 \sum_i\omega_i^2  \mathbb{R}_{iy}'^2 R_y^2-\sum_i\frac{\hbar^2{K_i}^2}{3M^2}-\frac{48Nc_0}{MR_x R_y R_z}  F_A\left( \frac{K_z}{K_x},\frac{K_z}{K_y}, \theta,\varphi,\theta'',\varphi''\right) \nonumber \\ 
 +\frac{48Nc_0}{MR_x R_y R_z} \left[F_A\left(\frac{R_x}{R_z},\frac{R_y}{R_z},\theta,\varphi,\theta',\varphi' \right)-R_y\partial_{R_y}F_A\left(\frac{R_x}{R_z},\frac{R_y}{R_z},\theta,\varphi,\theta',\varphi' \right)\right]
 &=0 \, , \label{VSTRy}\\ 
 \hspace*{-2cm}
\sum_i\omega_i^2  \mathbb{R}_{iz}'^2 R_z^2-\sum_i\frac{\hbar^2{K_i}^2}{3M^2}-\frac{48Nc_0}{MR_x R_y R_z} F_A\left( \frac{K_z}{K_x},\frac{K_z}{K_y}, \theta,\varphi,\theta'',\varphi''\right)
\nonumber \\ 
+\frac{48Nc_0}{MR_x R_y R_z} \left[F_A\left(\frac{R_x}{R_z},\frac{R_y}{R_z},\theta,\varphi,\theta',\varphi' \right)-R_z\partial_{R_z}F_A\left(\frac{R_x}{R_z},\frac{R_y}{R_z},\theta,\varphi,\theta',\varphi' \right) \right]
&=0 \,,\label{VSTRz}\\
 \sum_{i,j}\omega_i^2  \mathbb{R}_{ij}'\partial_{\theta'}\mathbb{R}_{ij}'R_j^{2}-\frac{48N c_0}{ MR_x R_y R_z}\partial_{\theta'}F_A\left(\frac{R_x}{R_z},\frac{R_y}{R_z},\theta,\varphi,\theta',\varphi' \right)&=0,\label{VSTT}\\
 \sum_{i,j}\omega_i^2  \mathbb{R}_{ij}'\partial_{\varphi'}\mathbb{R}_{ij}'R_j^{2}-\frac{48N c_0}{ MR_x R_y R_z}\partial_{\varphi'}F_A\left(\frac{R_x}{R_z},\frac{R_y}{R_z},\theta,\varphi,\theta',\varphi' \right)&=0 \, ,\label{VSTF}\\
  \partial_{\theta''}F_A\left(\frac{K_z}{K_x},\frac{K_z}{K_y},\theta,\varphi,\theta'',\varphi'' \right)&=0  \, ,\label{VSTTT}\\
 \partial_{\varphi''}F_A\left(\frac{K_z}{K_x},\frac{K_z}{K_y},\theta,\varphi,\theta'',\varphi'' \right)&=0 \, .\label{VSTFF}
\end{eqnarray}
From Eq.~(\ref{KxKy}) we see that the FS remains cylindrically symmetric even in the case of a general orientation of the dipoles with respect to the trap.

\section{Orientation of the Fermi surface}

Equations~(\ref{VSTTT}) and (\ref{VSTFF}) can be solved analytically, independently of other equations, yielding the physically expected result $\theta''=\theta$, $\varphi''=\varphi$. This means that the FS stretches along the dipoles' orientation, as it was verified both experimentally and theoretically for the atomic erbium gas \cite{Veljic2}. Here we obtain this result self-consistently within our approach, which demonstrates that ansatz (\ref{eq:ansatzwigner}) properly captures the ground-state properties of dipolar Fermi gases.

\section{Dimensionless form of equations for the ground state}

If we eliminate the angles $\theta'',\varphi''$ as outlined above and set $\theta''=\theta$, $\varphi''=\varphi$ in all equations, the system is now determined by the 8 variational parameters ($R_i, K_i, \theta',\varphi'$), which are obtained by solving the set of equations (\ref{VSTN})--(\ref{VSTF}). They can be transformed into a dimensionless form by expressing the TF radii $R_i$ and momenta $K_i$ in units of $R_i^0=\sqrt{\frac{2E_{\rm F}}{M \omega_i^2}}$ and $K_{\rm F}=\sqrt{\frac{2ME_{\rm F}}{\hbar^2}}$, respectively, where $E_{\rm F}=\hbar (6N\omega_x \omega_y \omega_z)^{1/3}$ stands for the Fermi energy. The quantities $R_i^0$ and $K_{\rm F}$ represent the TF radii and Fermi momentum of the ideal Fermi gas, illustrated in Fig.~\ref{fig:fig1}(a). The dimensionless radii and momenta are defined by $\tilde{R}_i=R_i/R_i^0$ and $\tilde{K}_i=K_i/K_{\rm F}$, and if we drop, for simplicity, the tilde signs, the set of equations (\ref{VSTN})-(\ref{VSTF}) reduces to
\begin{eqnarray} 
1-R_x R_y R_zK_x K_y K_z&=0\, ,\label{dVSTN} \\
{ K}_x-{ K}_y&=0\, , \label{dKxKy}  \\
2{ K}_x^2-{ K}_y^2-{ K}_z^2+\frac{3\varepsilon_{\rm dd} c_{\rm d}}{R_x R_y R_z}\frac{{K}_z}{{K}_z}\partial_{{K}_x} f\left( \frac{{ K}_z}{{ K}_x}, \frac{{ K}_z}{{ K}_y}\right)&=0 \, ,
\label{dcylindersymmglobalmoment}\\
 \hspace*{-2.2cm}
\sum_i\frac{\omega_i^2}{\omega_x^2} \mathbb{R}_{ix}'^2{R}_x^2-\frac{1}{3}\sum_i{ K}_i^2-\frac{\varepsilon_{\rm dd} c_{\rm d}}{R_x R_y R_z}f\left( \frac{{ K}_z}{{ K}_x}, \frac{{ K}_z}{{ K}_y}\right)\nonumber \\
+\frac{\varepsilon_{\rm dd} c_{\rm d}}{R_x R_y R_z}\left[F_A\left(\frac{{ R}_x \omega_z}{{ R}_z \omega_x},\frac{{ R}_y \omega_z}{{ R}_z \omega_y},\theta,\varphi,\theta',\varphi'\right) -{R}_x\partial_{{R}_x}F_A\left(\frac{{ R}_x \omega_z}{{ R}_z \omega_x},\frac{{ R}_y \omega_z}{{ R}_z \omega_y},\theta,\varphi,\theta',\varphi'\right)\right]&=0\, , \label{dVSTRx} \\
\sum_i\frac{\omega_i^2}{\omega_y^2} \mathbb{R}_{iy}'^2{R}_y^2-\frac{1}{3}\sum_i{ K}_i^2-\frac{\varepsilon_{\rm dd} c_{\rm d}}{R_x R_y R_z}f\left( \frac{{ K}_z}{{ K}_x}, \frac{{ K}_z}{{ K}_y}\right)\nonumber \\
+\frac{\varepsilon_{\rm dd} c_{\rm d}}{R_x R_y R_z} \left[F_A\left(\frac{{ R}_x \omega_z}{{ R}_z \omega_x},\frac{{ R}_y \omega_z}{{ R}_z \omega_y},\theta,\varphi,\theta',\varphi'\right) -{R}_y\partial_{{R}_y}F_A\left(\frac{{ R}_x \omega_z}{{ R}_z \omega_x},\frac{{ R}_y \omega_z}{{ R}_z \omega_y},\theta,\varphi,\theta',\varphi'\right) \right]
&=0\, , \label{dVSTRy} \\
\sum_i\frac{\omega_i^2}{\omega_z^2} \mathbb{R}_{iz}'^2{R}_z^2-\frac{1}{3}\sum_i{ K}_i^2-\frac{\varepsilon_{\rm dd} c_{\rm d}}{R_x R_y R_z} f\left( \frac{{ K}_z}{{ K}_x}, \frac{{ K}_z}{{ K}_y}\right)\nonumber \\
+\frac{\varepsilon_{\rm dd} c_{\rm d}}{R_x R_y R_z} \left[F_A\left(\frac{{ R}_x \omega_z}{{ R}_z \omega_x},\frac{{ R}_y \omega_z}{{ R}_z \omega_y},\theta,\varphi,\theta',\varphi'\right)-{R}_z\partial_{{R}_z}F_A\left(\frac{{ R}_x \omega_z}{{ R}_z \omega_x},\frac{{ R}_y \omega_z}{{ R}_z \omega_y},\theta,\varphi,\theta',\varphi'\right)\right]&=0\, , \label{dVSTRz} \\
\sum_{i,j}\frac{\omega_i^2}{\omega_j^2}  \mathbb{R}_{ij}'\partial_{\theta'}\mathbb{R}_{ij}'{R}_j^{2}-\frac{\varepsilon_{\rm dd} c_{\rm d}}{R_x R_y R_z} \partial_{\theta'}F_A\left(\frac{{ R}_x \omega_z}{{ R}_z \omega_x},\frac{{ R}_y \omega_z}{{ R}_z \omega_y},\theta,\varphi,\theta',\varphi' \right)&=0,\label{dVSTT}\\
 \sum_{i,j}\frac{\omega_i^2}{\omega_j^2}  \mathbb{R}_{ij}'\partial_{\varphi'}\mathbb{R}_{ij}'{R}_j^{2}-\frac{\varepsilon_{\rm dd} c_{\rm d}}{R_x R_y R_z} \partial_{\varphi'}F_A\left(\frac{{ R}_x \omega_z}{{ R}_z \omega_x},\frac{{ R}_y \omega_z}{{ R}_z \omega_y},\theta,\varphi,\theta',\varphi' \right)&=0 \, ,\label{dVSTF}
\end{eqnarray}
where $\varepsilon_\mathrm{dd}=\frac{d^2}{4\pi\varepsilon_0}\sqrt{\frac{M^3}{\hbar^5}}(\omega_x \omega_y \omega_z N)^{1/6}$ is the dimensionless relative DDI strength and $c_{\rm d}=\frac{2^{\frac{38}{3}}}{3^{\frac{23}{6}} \cdot 5 \cdot 7 \cdot \pi^2}$ is a number.

\section{Beyond-mean-field corrections}

Here we estimate beyond-mean-field effects in the calculation of the Fermi surface shape and the stability of the system for strong dipolar interaction. We follow Ref.~\cite{Kopietz1}, which derives beyond-mean-field corrections to both the Fermi surface deformation and compressibility of the system. Note that this reference considers a homogeneous system, and that the estimates based on these results might not be fully applicable to a trapped system. However, the corresponding results for a trapped system are not available, and therefore we use Ref.~\cite{Kopietz1} to estimate beyond-mean-field corrections in our case. In order to do so, we identify the homogeneous density of Ref.~\cite{Kopietz1} with the average density of the trapped system, calculated as $N/V$, where $V=\frac{4\pi}{3} R_x R_y R_z$ is a volume of the TF ellipsoid in real space for specific parameters in the experimental setup.

\begin{figure*}[!t]
\centering
\includegraphics[height=4.8cm]{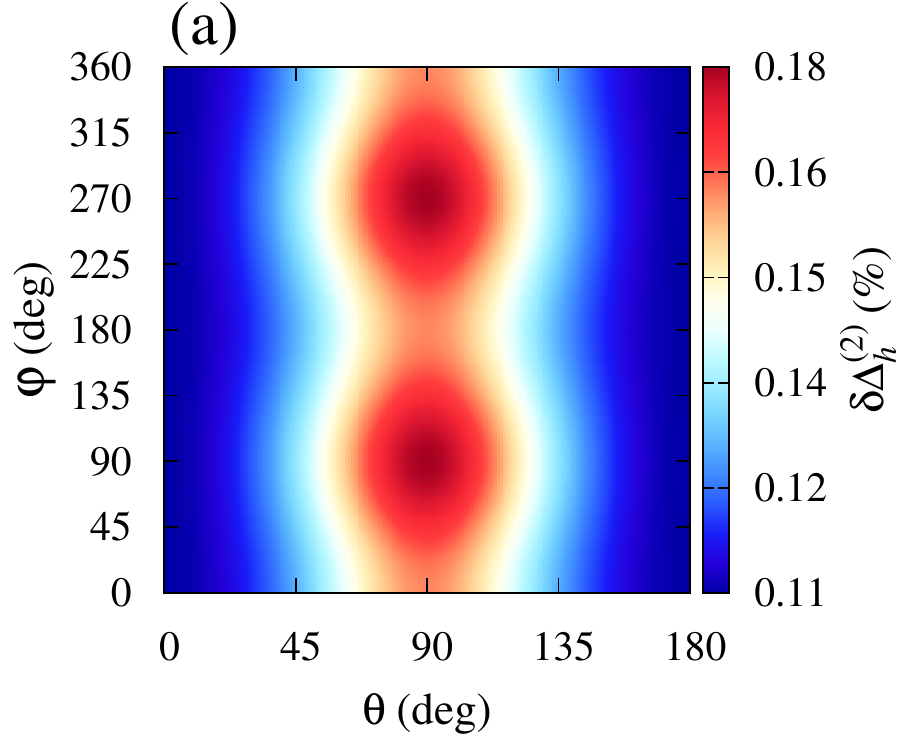}\hspace*{1cm}
\includegraphics[height=4.8cm]{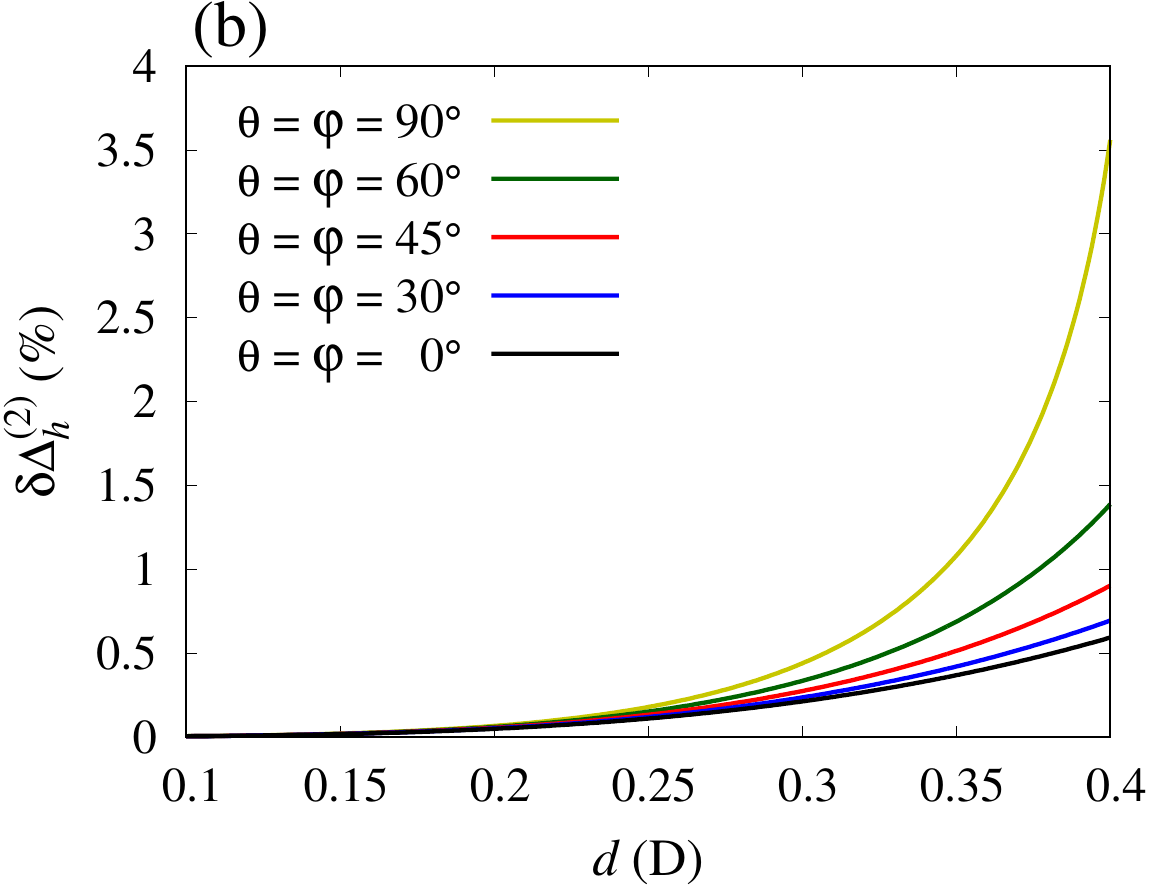}
\caption{Beyond-mean-field corrections to the FS deformation for the trap parameters $(\omega_x,\, \omega_y,\, \omega_z)=2\pi\times (63,\, 36,\, 200)$~Hz of Ref.~\cite{KRb-qd}, with $N=3\cdot 10^4$ molecules: (a) angular dependence for $d=0.25$~D, which is used to obtain Fig.~\ref{fig:fig4}(b); (b) the corresponding dependence on the dipole moment $d$ for fixed values of tilt angles $\theta$ and $\varphi$.}
\label{fig:figS1}
\end{figure*}

Having this in mind, we first use Eq.~(44) from Ref.~\cite{Kopietz1} to estimate beyond-mean-field corrections to the Fermi surface deformation. This is illustrated in Fig.~\ref{fig:figS1}(a) for experimental system parameters \cite{KRb-qd} with $d=0.25$~D, which are used to obtain Fig.~\ref{fig:fig4}(b). It turns out that corrections are just a fraction of one percent. In Fig.~\ref{fig:figS1}(b) we see how the beyond-mean-field correction depends on the dipole moment. It amounts to a few percent even for the strongest values of $d$ that can be achieved in current experiments with $^{40}$K$^{87}$Rb \cite{KRb-qd}.

\begin{figure*}[!t]
\centering
\includegraphics[height=4.8cm]{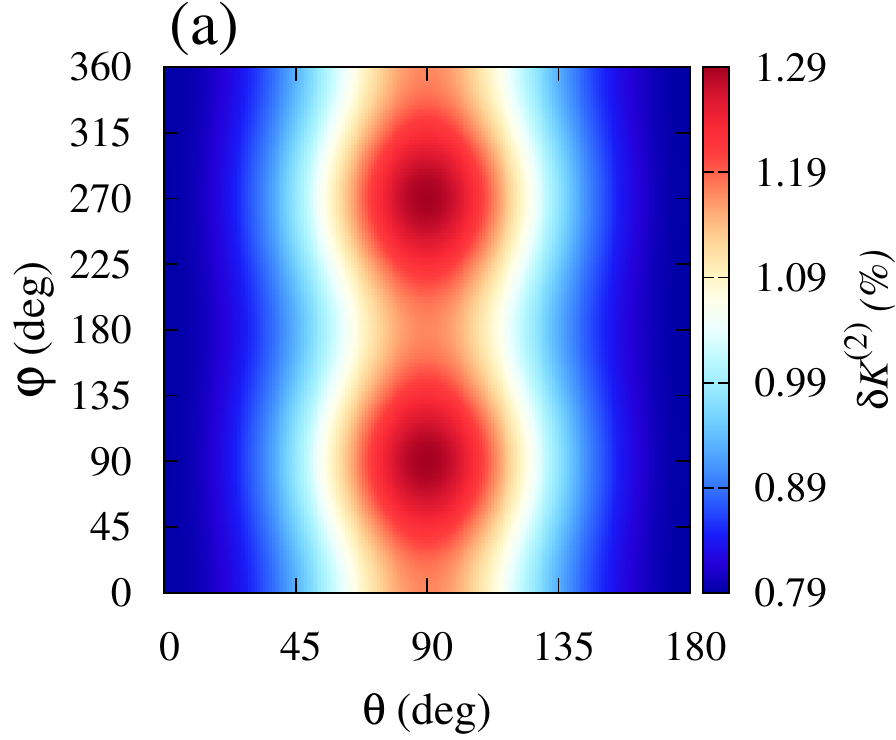}\hspace*{1cm}
\includegraphics[height=4.8cm]{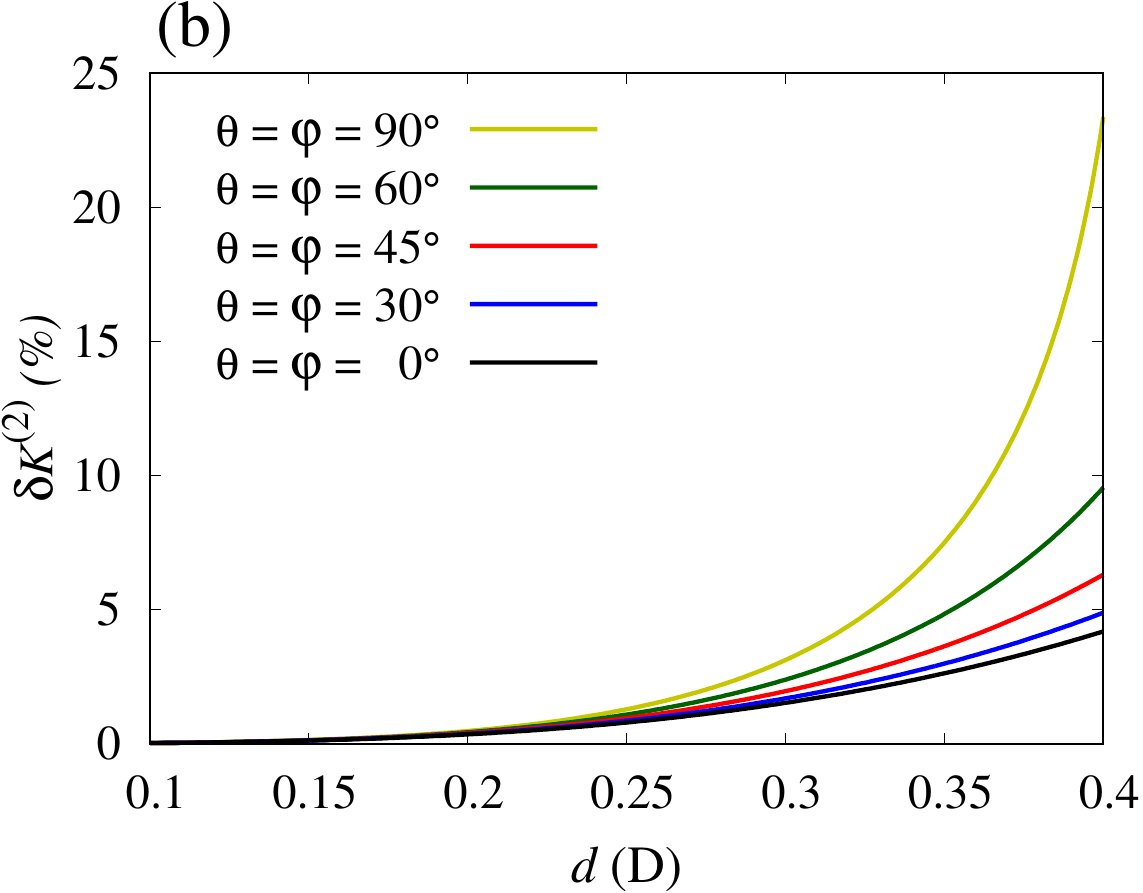}
\caption{Beyond-mean-field corrections to the system's inverse compressibility  $\delta K^{(2)}$ for trap parameters $(\omega_x,\, \omega_y,\, \omega_z)=2\pi\times (63,\, 36,\, 200)$~Hz of Ref.~\cite{KRb-qd}, with $N=3\cdot 10^4$ molecules: (a) angular dependence for $d=0.25$~D, corresponding to Fig.~\ref{fig:fig4}(b); (b) $\delta K^{(2)}$ as a function of the dipole moment $d$ for fixed values of tilt angles $\theta$ and $\varphi$.}
\label{fig:figS2}
\end{figure*}

However, the situation is more complex when we consider the bulk modulus, i.e., the inverse compressibility $K$ of the system, which is used to estimate the stability border according to the Pomeranchuk criterion \cite{pom1, pom2}, and whose beyond-mean-field correction is given by Eq.~(46) in Ref.~\cite{Kopietz1}. For instance, for system parameters used to obtain Fig.~\ref{fig:fig4}(b), the corresponding second-order correction to the inverse compressibility $\delta K^{(2)}$ is of the order of one percent, as can be seen in Fig.~\ref{fig:figS2}(a), where we plot its angular dependence. These corrections are calculated for the dipole moment value $d = 0.25$~D. In Fig.~\ref{fig:figS2}(b) we see, however, that the correction can be much higher for larger values of $d$, and that it strongly depends on the orientation of the dipoles. If we use a 10\% threshold for the inverse compressibility correction, we see that $d$ can be as high as 0.35~D in the worst-case scenario, when the dipoles lie within the pancake plane, while for other values of the angles one can use even larger values of $d$. Taking into account that this coincides with the maximal achievable dipole moment in the current experiment with $^{40}$K$^{87}$Rb \cite{KRb-qd}, for their trap configuration our mean-field theory is applicable with reasonable accuracy, as shown in Fig.~\ref{fig:figS2}. However, for other trap configurations the mean-field theory could break down for smaller values of $d$, as the inverse compressibility $K$ of the system can have a strong angular dependency, depending on the underlying trap geometry. One can use a similar calculation as the one presented here to make an appropriate estimate for any given trap configuration.

\section{Comparison of stability diagrams obtained with self-consistent and fixed cloud orientation}

\begin{figure*}[!ht]
\centering
\includegraphics[height=4.8cm]{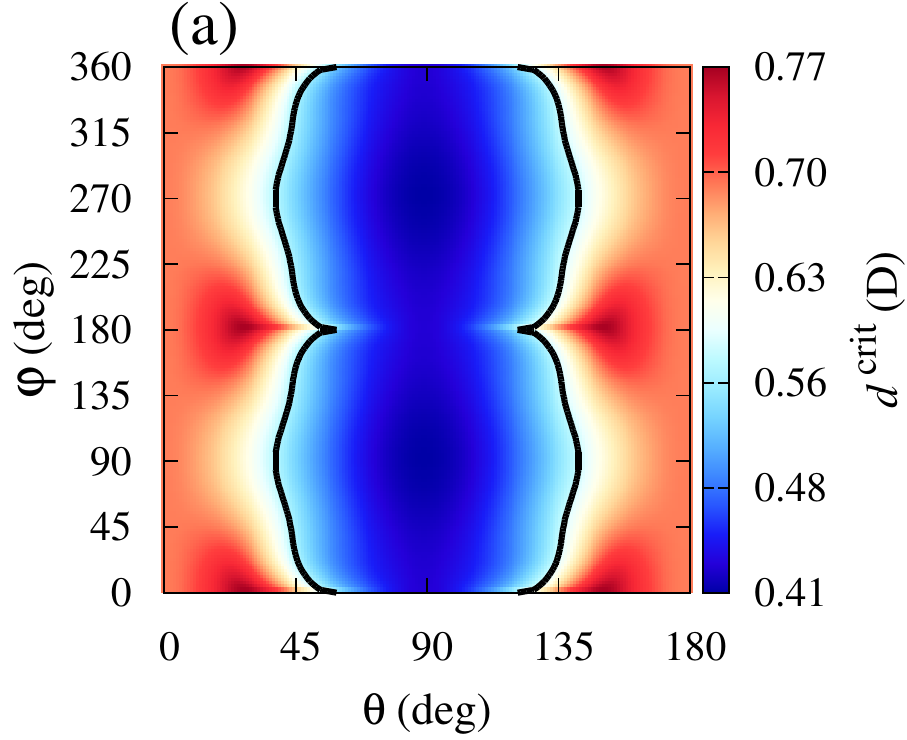}\hspace*{1cm}
\includegraphics[height=4.8cm]{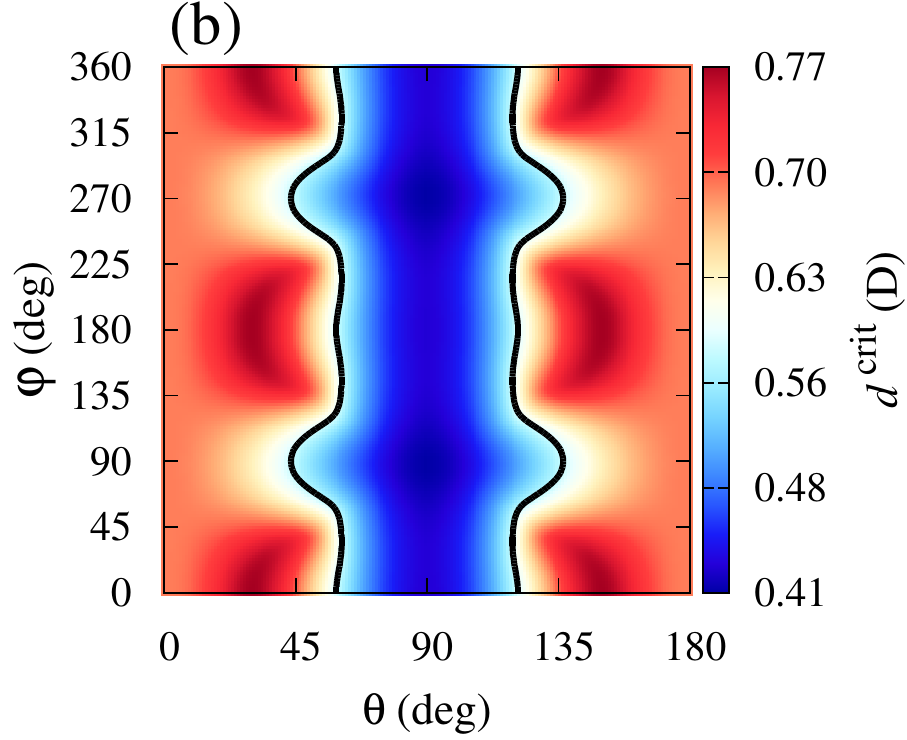}
\caption{Angular stability diagram of $^{40}$K$^{87}$Rb obtained by: (a) solving a full set of equations (\ref{dVSTN})--(\ref{dVSTF}), with $\theta', \varphi'$ treated as free variational parameters; (b) assuming $\theta'=\varphi'=0$ and solving a reduced set of equations (\ref{dVSTN})--(\ref{dVSTRz}). Black line corresponds to the permanent electric dipole moment $d=0.574$~D of $^{40}$K$^{87}$Rb. The trap frequencies are as in Ref.~\cite{KRb-qd} and $N=3\cdot 10^4$.}
\label{fig:figS3}
\end{figure*}

The orientation of the dipoles with respect to the harmonic trap affects not only the shape of the molecular cloud in real space, but also its orientation. While previously it was always assumed that the axes of the molecular cloud coincide with the axes of the trap ($\theta'=\varphi'=0$), our theory takes into account the effects of the DDI and determines the angles $\theta',\varphi'$ in a self-consistent manner. In Fig.~\ref{fig:figS3} we compare stability diagrams, expressed in terms of the critical electric dipole moment $d^\mathrm{crit}$, obtained in panel (a) by our theory and in panel (b) by assuming $\theta'=\varphi'=0$. The numerically calculated angular distributions are markedly different and the stability region is reduced when the full theory is applied. This makes the approach presented here important for the design of new experiments with polar molecules, in particular in the strong DDI regime.

\end{document}